\documentclass{aa}
\usepackage{graphics}

\begin{document}

   \thesaurus{23     
              (04.01.1;  
               04.03.1)}  
   \title{The SIMBAD astronomical database}

   \subtitle{The CDS Reference Database for Astronomical Objects}

\author{
    Marc Wenger
\and
   Fran\c{c}ois Ochsenbein
\and
    Daniel Egret
\and
    Pascal Dubois
\and
    Fran\c{c}ois Bonnarel
\and
    Suzanne Borde\thanks{DASGAL, Observatoire de Paris}
\and
   Fran\c{c}oise Genova
\and
   G\'erard Jasniewicz\thanks{Groupe de Recherche
   en Astronomie et Astrophysique du Languedoc (GRAAL), Montpellier}
\and
   Suzanne Lalo\"e
\and
   Soizick Lesteven
\and
   Richard Monier
          }

   \offprints{Daniel Egret}
   \mail{question@simbad.u-strasbg.fr}

   \institute{CDS, Observatoire astronomique de Strasbourg, UMR 7550,
11 rue de l'Universit\'e, F-67000 Strasbourg, France}

   \date{Received 6 December 1999 / Accepted 16 December 1999}

\maketitle

   \begin{abstract}
{\sc Simbad} is the reference database for identification and
bibliography of astronomical objects. It contains identifications,
`basic data', bibliography, and selected observational
measurements for several
million astronomical objects. 

{\sc Simbad} is developed and maintained by CDS, Strasbourg.
Building the database contents is achieved
with the help of several contributing institutes. Scanning
the bibliography is the
result of the collaboration of CDS with bibliographers
in 
Observatoire de Paris (DASGAL), 
Institut d'Astrophysique de Paris,
and Observatoire de Bordeaux.

When selecting catalogues and tables for inclusion,
priority is given to optimal
multi-wavelength coverage  of the database, and to
support of research developments 
linked to large projects.
In parallel, the systematic scanning of the
bibliography reflects the diversity and general 
trends of astronomical research.

    A  WWW interface to {\sc Simbad} is available at:

{http://simbad.u-strasbg.fr/Simbad}.


\end{abstract}

%

\section{Introduction}

\subsection{The CDS}
The Centre de Donn\'ees astronomiques de Strasbourg
(CDS) defines, develops, and maintains services 
to help the astronomers find the information
they need from the
very rapidly increasing wealth of astronomical information, and
particularly of on-line information. 

CDS is operated at the Strasbourg astronomical
Observatory, under an agreement between French
Institut National des Sciences de l'Univers (INSU) and
Universit\'e Louis Pasteur, Strasbourg (ULP). CDS personnel
created and implemented the {\sc Simbad}  data bank and maintain its data
and software system. 

A detailed description of the CDS on-line services can be found, e.g., 
in Egret et al.\ (\cite{cds-amp2})
and in Genova et al.\ (\cite{cds-hub}, \cite{cds}, \cite{cds2000}), 
or at the CDS web site\footnote{{\em Internet address:}
http://cdsweb.u-strasbg.fr/}. 
 Questions or comments about the
CDS services can be sent to the hot line {\it
question@simbad.u-strasbg.fr}.

\subsection{SIMBAD}
 
The {\sc Simbad} database contains identifications,
`basic data', bibliographical references,
and selected observational measurements for more than
2.7 million astronomical objects (November 1999).
Data and information published in {\sc Simbad}
come from selected catalogues and tables
and from the whole astronomical literature.

The specificity of the {\sc Simbad} database is to
organize the information per astronomical object,
thus offering a unique perspective on astronomical data.
This is done through a careful cross-identification 
of objects from catalogues, lists, and journal articles.
The ability to gather together any sort of published
observational data  related to stars or galaxies has 
made {\sc Simbad} a key tool used worldwide for all kinds 
of astronomical studies.

{\sc Simbad} is the acronym for  {\sl S}et of
                            {\sl I}dentifications, 
                            {\sl M}easurements
            and             {\sl B}ibliography
            for             {\sl A}stronomical
                            {\sl D}ata.
 
The main access point to {\sc Simbad} 
is the WWW home page\footnote{http://simbad.u-strasbg.fr/Simbad};
there is a mirror copy at SAO, 
Harvard\footnote{http://simbad.harvard.edu/Simbad}.

\subsection{Historical background}

Building a reference database for stars  -- and, later, for
extragalactic objects and all astronomical objects outside
the Solar System -- has been the first goal of the CDS: {\sc Simbad}
is the result of an on-going effort which started soon after
the creation of CDS in 1972.
{\sc Simbad} was created by merging the Catalog of Stellar Identifications
(CSI, Ochsenbein et al.\ \cite{csi}) and the Bibliographic Star Index 
(Ochsenbein \cite{bsi})
as they existed until 1979.  The resulting data base 
(at that time, about 400,000 objects, mainly stars) was then
expanded by the addition of source data from the catalogs and
tables, and by  new literature references.  
The database was extended to galaxies
and other non-stellar objects in 1983 (Dubois et al. \cite{dubois83}).
For details about the early developments of {\sc Simbad} see
Egret (\cite{story}).
 
The first on-line interactive version
of {\sc Simbad} was released in 1981, and operated 
at  the Strasbourg Cronenbourg computer
center until December 1984, 
when it was moved to Universit\'e Paris-Sud at
Orsay, and operated there until June 30, 1990. 
The database is now hosted on a 
Unix server, at the Strasbourg Observatory. 
  
The original command line interface has been complemented by
an interactive X-Window interface ({\sc XSimbad}) in 1994,
and by a World-Wide Web interface in 1996.
There is also a client/server mode,  providing quick responses
to simple queries, essentially for the name resolution in
archives and information systems (see Section~\ref{interface}).

For descriptions of earlier stages of the database, 
see Heck \& Egret (\cite{messenger}), and Egret et al.\
(\cite{ampersand}).

\section{SIMBAD main features}

 {\sc Simbad} is, in the first place, a database of identifications,
aliases and names of astronomical objects: in principle any name
found in the literature -- provided it is given as a
syntactically  correct character string -- can be submitted
to {\sc Simbad} in order to
retrieve basic information known for this object,
as well as pointers to complementary data and bibliography.
This implies a continuous careful cross-identification 
of objects from catalogues, lists, and journal articles.
This ability to gather together any sort of published
observational data  related to stars or galaxies is
the first key feature of {\sc Simbad}.

Another unique feature is the complete and up-to-date
bibliographic survey of the astronomical literature:
objects are associated with the references of all
papers in which they are mentioned, independently of the
aliases used to name the object.

In addition, the Dictionary of Nomenclature 
(Section \ref{Dic}) 
is an essential tool for managing the very complex nomenclature
of objects found in the literature, and for matching
naming variations with those adopted or simply accepted by
{\sc Simbad}. It also includes hints for helping to solve ambiguities, 
according to the type of object, or to the format.
This is complemented by the {\em sesame} module 
within {\sc Simbad},
for the management of possible variations in the
naming of astronomical objects.

The database management system of {\sc Simbad}
(Section \ref{dbms}) has been developed 
in-house at CDS, using
the concepts of object-oriented programming. 

{\sc Simbad} is kept up-to-date (Section \ref{update})
on a daily basis, as the
result of the collaboration of CDS with bibliographers
in Institut d'Astrophysique de Paris and the
Paris and Bordeaux observatories.

The statistical contents of the database 
(Section~\ref{stats}) can be summarized 
in a few figures as follows
(the figures quoted are statistics of November 1999): 
\begin{itemize}
\item 	entries for about 2.7 million astronomical 
objects (stars, galaxies and all astronomical objects 
outside the solar system);
\item 	a cross-index of 7.5 million identifiers
related to 4500 astronomical catalogues and tables,
lists, and observation logs of space missions;
\item 	observational data from some 25 different 
types of data catalogues and compilations;
\item  a bibliographic survey covering the astronomical 
literature since 1950 for stars, and since 1983 for 
extragalactic objects: more than 3 million citations from
110,000 different papers.
\end{itemize}

\section{SIMBAD astronomical contents}
\label{stats}

\subsection{The objects}
 
The {\sc Simbad} data base presently contains information for about:
 
\begin{itemize}
 \item  1,500,000 stars; 
 \item    450,000 galaxies;
 \item    100,000 other non-stellar objects (planetary nebulae, 
       clusters, \ion{H}{ii} regions, etc.);
 \item    and some 650,000 additional objects observed at various
          wavelengths (Radio, IR, X), and for which 
          classification is not yet assigned.
\end{itemize}
 
The only astronomical objects
specifically excluded from {\sc Simbad} are the Sun and
Solar System bodies.

The {\sc Simbad} database is primarily organized per astronomical object.
The aim is to provide, as much as possible, the user 
with all published information 
(identifications, observational
data, bibliographical references, and pointers towards
external archives) concerning any given object, or list 
of objects.

The two main channels for feeding the database are the following:
\begin{itemize}
\item  the daily scanning of papers published in the astronomical 
literature provides new references and new identifiers for existing objects, 
as well as opportunities to create new objects,  
using the basic data possibly given in the article;
\item the complete (or partial) folding of selected catalogues 
into the database serves as a basis for improving the 
completeness and multi-wavelength coverage of the database.
 \end{itemize}

Catalogues are selected for integration with
priority given to those which can
help provide an optimal support for 
the large projects conducted within the astronomical community.
A large effort was, for instance, devoted in recent years to stellar
catalogues (PPM, HIC, CCDM), in the context of the Hipparcos project,
and to multi-wavelength identifications (IRAS PSC, Einstein 1E and
2E catalogues, older X--ray catalogues, the IUE Merged Log, etc.). 
The Hipparcos and Tycho catalogues (ESA \cite{tyc})
have been recently included, and inclusion of the ROSAT All
Sky Survey is planned in the near future.

In parallel, the systematic scanning of the
bibliography reflects the diversity and general trends of research
in astronomy, and takes into account shorter lists. The published lists
from the microlensing surveys, or e.g. the EUVE catalogues, were folded
into the database as a result of this scanning.

When an object is
found in the literature or in a catalogue, its possible
cross-identification with objects already in {\sc Simbad} is
systematically studied, before entering the reference 
and the new object name in the database. 
About 4500 different names of
catalogues or object lists from published papers can 
currently be found in {\sc Simbad},
covering all the wavelength domains from high energy astrophysics to
radio. 

When no proper name is suggested by the authors, or 
when the acronym generates an ambiguity with already existing ones, 
the current practice, shared with the NED database, is to create
an acronym within brackets using the initials of the last names
of the first three authors, and the year of publication.
For example, [HFE83] 366 is the 366th entry in the main table 
of a paper by Helmer, Fabricius, Einicke and colleagues published in 1983.
From the year 2000 on, the year will be noted with four digits (e.g.,
[ABC2000]).

Many objects have more than one name, since the database
contains more than 7.5 million object names for 
2.7 million objects. 
Examples of objects with more than 50 identifiers,
are the galaxy \object{M 87} in Virgo, the bright stars 
\object{Procyon} and \object{Capella},
the quasar \object{3C 273}, the \object{Crab Nebula}.

To help the users with the complex nomenclature
of astronomical objects, the CDS now maintains
 and distributes on--line
the Dictionary of Nomenclature of Celestial Objects 
(first developed by Lortet et al.\   \cite{dic2}; see 
Section~\ref{Dic}).

\subsection{The Data}

 In the following, the word {\em object\/} will be used
to designate a star, non stellar object, or collection of
objects such as a cluster, which corresponds to an individual
entry in {\sc Simbad}.  
For each object, the following data are included when available:
 
\begin{itemize}
 
\item   Basic data:
 
 \begin{description}
 \item[stars]: object type, coordinates, proper motion, 
     $B$ and $V$ magnitudes, spectral type, parallax, radial velocity;
  \item[galaxies]:  object type, coordinates, blue and visual 
          integrated magnitudes,
            morphological type, dimension, radial velocity or redshift.
  \item[other]: object type, position, $B$ and $V$ magnitudes.
\end{description}
 
 \item  Cross-identifications from some 4500 catalogues 
and tables, either
completely or partially included in the data base.  
 
 \item  Observational data (also called {\em
measurements\/}), for some 25  data types. 
A list of these types is given in the Table~\ref{measurements}.

 \item  General bibliography,
including references to all published papers since 1983
citing the object under any of its designations.
For stars, the bibliography starts as early as 1950, but
with a smaller coverage of the literature. 
{\sc Simbad} also includes a few hundred references before 1950,
but without any systematic trend.
The bibliography gives
access to abstracts and electronic articles when
available (either directly from publishers, or through ADS).
Currently about 100 journals
covering the complete astronomical literature are 
regularly scanned.
A complete list is available 
on line\footnote{http://simbad.u-strasbg.fr/guide/chH.htx}.
 
\end{itemize}

In the following, a more detailed description of
some of these elements is given.

\subsubsection{Object type}

The object type refers to a
hierarchical classification of the objects in {\sc Simbad},
derived by the CDS team on the basis of
the catalogue identifiers (as proposed by Ochsenbein
and Dubois \cite{type}).
From {\em Star}  to  {\em  Maser source}, or {\em Cluster of Galaxies},
some 70 different categories, general, or very specific, are
proposed  (see examples in Table~\ref{object-type}).
A complete list is available 
on line\footnote{http://simbad.u-strasbg.fr/guide/chF.htx}.

This classification is intended to help the user select objects
out of the database
(e.g.\ through the filter procedure, see Section~\ref{filter}). 
It is also a powerful tool for data cross--checking
and quality control.
It has been designed to be practical and useful,
and complements other features also available in {\sc Simbad} 
(morphological type or spectral type information, catalogues, and
measurements).
It can follow the evolution of astronomy,
with the introduction of new categories recently
appeared in the literature (e.g., in the last years,
Low-Mass or High-Mass X-Ray binary, Microlensing event, or
Void).

Each class has normally a standard designation, a condensed one (used in
    tables) and an extended explanation.
The classification uses a hierarchy with four levels, reflecting
our knowledge of the characteristics of the astronomical object.
For instance, an object can be classified as a ``Star'' (this is
level 1). If photometric observations have shown variability
of the object, it can be classified as a ``Variable star'' (this
is level 2). Examples of level 3 and 4 are ``Pulsating variable'',
and ``Cepheid''.

This hierarchy of object types (and their possible synonyms) 
is managed
in the database in such a way that selecting variable stars 
({\tt V*})
is understood as selecting objects classified
as {\tt V*}, and all subdivisions (e.g. {\tt PulsV*}, {\tt Mira},
{\tt Cepheid}, etc.). 
If the user is only interested in RR Lyrae type stars, he/she will
use the {\tt RRLyr} type, leaving aside all other variable stars
for which the variability mode is different, or not known.

The classification emphasizes the physical nature of the object
rather  than a peculiar emission in some region of the
electromagnetic spectrum or the location in peculiar 
clusters or external galaxies.
Therefore objects  are classified as peculiar emitters 
in a given wavelength (such as UV or IR source)
only if nothing more about the nature of the object is known, 
i.e. it cannot be decided on the sole basis
of the basic data whether the object is a star, 
a multiple system, a nebula, or a galaxy. 
For instance, if an object appears only in the
IRAS catalogue, it is automatically
classified as IR object: it is left
to the user to decide to go further
and to derive, e.g. on the basis of the
IRAS colors, the probability for the source to be stellar
or extragalactic.

\begin{table}
\caption{Object type classification: extracts from the object type table
  illustrating examples of the four levels of the classification scheme}
\footnotesize
\begin{tabular}{lllp{4cm}}
\hline 
 Level  &   Standard    &   Short  & Extended Explanation \\
    &          name     &  name    &    \\
\hline
 \dots \\
1 &   Star               & *        & Star \\ 
2 &     *inCl            & *iC      & Star in Cluster \\
2 &     *inNeb           & *iN      & Star in Nebula \\
2 &     *inAssoc         & *iA      & Star in Association \\
2 &     *in**            & *i*      & Star in double system \\
2 &     V*?              & V*?      & Star suspected of Variability \\
2 &     Pec*             & Pe*      & Peculiar Star \\
3 &       HB*            & HB*      & Horizontal Branch Star \\
3 &       YSO            & Y*O      & Young Stellar Object \\
3 &       Em*            & Em*      & Emission-line Star \\
4 &        Be*           & Be*      & Be Star \\
  \dots \\
1 &   Galaxy             & G        & Galaxy \\
2 &     PartofG          & PoG      & Part of a Galaxy \\
2 &     GinCl            & GiC      & Galaxy in Cluster of Galaxies \\
2 &     GinGroup         & GiG      & Galaxy in Group of Galaxies \\
2 &     GinPair          & GiP      & Galaxy in Pair of Galaxies \\
2 &     High\_z\_G       & HzG      & Galaxy with high redshift \\
  \dots \\
2 &     AGN              & AGN      & Active Galaxy Nucleus \\
3 &       LINER          & LIN      & LINER-type Active Galaxy Nucleus \\
3 &       Seyfert        & SyG      & Seyfert Galaxy \\
4 &         Seyfert\_1   & Sy1      & Seyfert 1 Galaxy \\
4 &         Seyfert\_2   & Sy2      & Seyfert 2 Galaxy \\
3 &       Blazar         & Bla      & Blazar \\
4 &         BLLac        & BLL      & BL Lac - type object \\
4 &         OVV          & OVV      & Optically Violently Variable \\
3 &       QSO            & QSO      & Quasar \\
\hline
\end{tabular}
\label{object-type}
\end{table}

Because there is at most one object type per object, this
classification should be used with caution when 
extracting samples out of the database.
This is typically the case for the wavelength types: using IR
or X as a criterion cannot generate a sample of all IRAS
sources, or all X-ray emitting objects, since a number 
of them are in fact classified as stars, galaxies, etc.

\subsubsection{Coordinates, Proper motion, Parallax,
     and Radial Velocity
             or Redshift}

The coordinates were originally stored in the database
in the FK4 system for equinox and epoch 1950.0. 
A major change was undergone in 1999, 
when they were moved to the 
International Celestial Reference System (ICRS, see
Feissel \& Mignard \cite{ICRS}) at epoch 2000.0, after the publication
of the Hipparcos and Tycho catalogues. 
The position data frame has become more complex, grouping together all
data needed for computing the coordinates into any reference frame, 
at any epoch and equinox: the coordinates themselves, the
proper motion,  the parallax and the radial velocity or
redshift.

All these data contain the same subfields: the original data,
displayed with a number of digits consistent with the announced
precision of the data; a quality code from 'A' (reference data)
to 'E' (unreliable origin); an error box (either a standard
error, or an ellipse), and the bibliographic reference of the
data.

In earlier versions of {\sc Simbad}, the determination of the position for
another equinox used to take only precession into account.
In the current version, a change of equinox takes into account not only 
the precession but also the proper motion, the reference frame (FK4, FK5, 
ICRS), and, when they are known, the parallax and radial
velocity.  When no epoch is specified, the year of the equinox
is used by default.

Data come from various sources. When astrometric data are 
available, the most accurate
one has been selected for the 'basic data'. 
Other values may 
be available as measurements (in the {\tt pos} type). The Hipparcos and
Tycho catalogues (ESA \cite{tyc}) constitute the major source of positions for
stars.

The coordinates precision may vary from $1\degr$ to $1/10$ mas. 
The default display format provides equatorial coordinates in the ICRS
system at epoch 2000.0, and in the FK5 system at equinoxes 2000 and
1950, as well as galactic coordinates. Coordinates in the FK4 system,
and ecliptic or super-galactic coordinates  can be computed on request.

The proper motions ($\mu_\alpha \cos\delta, \mu_\delta$) are given in
mas/year, together with their standard errors (in mas/year).
The primary source of proper motions is the Hipparcos and Tycho
catalogues (ESA \cite{tyc}).

The errors for positions or proper motions 
are expressed as error ellipses, made of three
numbers, within brackets: the major axis, the minor axis,
and the position angle of the major axis
(measured from North to East).
Major and minor axes are expressed in mas for the position,
and mas/yr for the proper motion;
the position angle is expressed in degrees,
in the range $[0\degr,180\degr[$.

When available, the stellar parallax is given in mas,
together with the associated error within brackets.
The primary source is the Hipparcos
and Tycho catalogues (ESA \cite{tyc}).

Radial velocity (in km/s), or redshift (for extragalactic objects)
are currently available for some 160,000 objects.
They are stored in their original type (either redshift,
or radial velocity in km/sec), associated with the standard error. 
Display can be done in the original type or forced to be one of the
two types, using the corresponding translation formula.

Stellar radial velocity data
have been compiled with the collaboration
of Observatoire de Marseille.

For extragalactic objects, up-to-date redshift information has 
recently been imported from the NASA/IPAC Extragalactic Database
(NED, Helou et al. \cite{ned}) as a result
of the ongoing  exchange agreement:
the {\sc Simbad} team is providing NED with bibliographic coverage 
of extragalactic objects for all astronomical journals, 
and is being given access, in return, to extragalactic 
data collected by NED.

Tables from individual articles constitute
the other major source of information.

\subsubsection{Magnitudes}

$B$ and $V$ magnitudes are given, when
possible,  in the Johnson's $UBV$ system. Both magnitudes
may be followed by a semicolon meaning they cannot be
made homogeneous to the $UBV$ system.  
In addition the following flags may appear:  
\begin{itemize} 
\item a `D' flags a joint magnitude in a double or
multiple system; 
\item a `V' indicates a variable magnitude and is
followed by a coded index giving a rough estimate of the
amplitude:

             $$\begin{tabular}{c l}   \hline
             code & definition \\   \hline
               1  &   1/100   mag.  \\
               2  &   1/10    mag.    \\
               3  &   1       mag. \\
               4  &   more than 1 mag.   \\
               ?  &   suspected variable   \\   \hline
             \end{tabular}$$
\end{itemize}

When possible the magnitudes have been taken from the
Tycho Reference Catalogue (H{\o}g et al.\ \cite{TRC})
where $B$ and $V$ magnitudes are derived from
the original $B_T$ and $V_T$.
Another major source is the
$UBV$ compilation of Mermilliod (\cite{UBV}).  
Otherwise the data would come from one of the published papers
associated to the object.

\subsubsection{Stellar Spectral type}

The spectral types of stars have been selected 
preferably in the
Michigan Catalogues of Two-Dimensional Spectral Types for
the HD stars  (Houk \cite{mss}, and seq.), or in the
bibliographical surveys of MK classifications (Jaschek
\cite{MK-MJ}).
In the absence of a full MK classification, the HD
spectral type is recorded.

 Most of the spectral types need
less than 5 characters, but this field can be as long as 36
characters.

\subsubsection{Morphological type and Dimension of galaxy}

The morphological types of galaxies have been 
selected primarily
from the Uppsala General Catalogue of Galaxies (UGC, Nilson \cite{ugc}), 
the Morphological Catalogue of Galaxies
(MCG, Vorontsov-Velyaminov, \cite{mcg}), and other
catalogues (see Dubois et al.\ \cite{dubois83}).

In complement, the following data, primarily from UGC,
are given, when available, for
galaxies:

$$\begin{tabular}{lp{6cm}}
$ \log D_{25}$   &   logarithm of the major axis $a$
expressed in tenths of arc minutes;  \\
$ \log R_{25}$   &   logarithm of the ratio $a/b$
where $a$ and $b$ are the major and minor axis;   \\
orientation      &   orientation angle (in degrees)  \\
(inclination)      &   inclination (in units of $15\degr$
from 0 to 7)  \\
\end{tabular}$$

\subsection{Cross--identifications}

\subsubsection{Aliases}

Cross--identifications of stars and galaxies have been
searched for {\sc Simbad} entries from (currently) about 4500 source
catalogues and tables, included, either completely or partially, in
the data base.
The index of 7.5 million {\em aliases\/}, thus constituted, is one of
the unique features of the {\sc Simbad} database.

Aliases may serve as entry points for related catalogues
and tables (e.g. in {\sc VizieR}). 
Cross-fertilization of a given research with
previous studies of the same object published in the astronomical
literature is made directly possible from the alias list.

The index of names and aliases constitutes the basis for the
{\sc Simbad} name resolver which provides, in response to any
object name,  the set of coordinates corresponding to the 
object position on the celestial sphere,
or the list of papers citing the object.
The name resolving power of {\sc Simbad} is used by many archives and
information systems (such as 
the archives of Hubble
Space Telescope or European Southern Observatory, 
the High Energy Astrophysics Science Archive Center, 
the Astrophysics Data System, 
servers of the Digitized Sky Surveys, etc.).

There is no {\sc Simbad} preferred name for objects\footnote{In the early times
of the  \emph{Catalog of Stellar Identifications} (Ochsenbein et al.\
\cite{csi}), the \emph{Durchmusterung} number had been used as a preferred
name for stars.}: all aliases can be equally used.  
A short list of major
catalogues is used internally to put at the top of the list 
the most common name according to the object type
(e.g., Messier or NGC identifier for galaxies and nebulae). 
All other identifiers are presented in alphabetical order.

A command of the {\sc Simbad} native node
(`{\tt selectid}'), and an option in
the sampling form of the WWW interface, 
allow the user to impose a list of
identifiers to be used when displaying object lists.

\subsubsection{Multiple systems}

It is to be noted that for a double system
in which the components can be observed separately,
{\sc Simbad} frequently includes three entries:  A
and B components, and an additional entry for 
the joint system (AB), the latter entry
carrying the observational data and references related
to the system as a whole.
This has to be taken into account in statistical studies
such as stellar counts.

\subsection{Observational data}

Observational data  are presently given for
the  measurement types listed in 
Table~\ref{measurements}.

\begin{table}
\caption{List of measurement types currently included in 
{\sc Simbad}, together with the number of entries
(October 1999).}
\footnotesize
\begin{tabular}{lp{5.5cm}r}
\hline 
Name     & Observational data           &  \multicolumn{1}{c}{\#}  \\
\hline 
{\tt CEL}   &Ultraviolet photometry from {\em Celescope}           &  5230 \\
{\tt Cl.G}  &Clusters of Galaxies  (Abell \& Corwin \cite{abcg})   &  5345 \\
  Einstein  &Einstein Observatory Soft X-ray Source List           &  5668 \\
{\tt GEN}   &$U B V B_1 B_2 V_1 G$ Geneva photometry               & 3650 \\ 
{\tt GJ}    &Absolute magnitudes and spatial velocities of nearby stars   & 2368 \\ 
{\tt Hbet}  &$H_\beta$ index                                       & 32278 \\ 
{\tt HGAM}  &$H_\gamma$ equivalent width                           &   723 \\
{\tt IRAS}  &IRAS Point Source Catalog                             &245784 \\
{\tt IRC}   &$KI$ photometry from {\em Two Micron Sky Survey}        &  4880 \\
{\tt IUE}   &International Ultraviolet Explorer (Merged Observation Log)& 66805 \\
{\tt JP11}  &$UBVRIJKLMNH$ 11-colour Johnson photometry            &  5892 \\
{\tt MK}    &Stellar spectral classification in Morgan-Keenan system &190231  \\
{\tt oRV}   &Stellar Radial velocities (also under {\tt GCRV})     & 68783 \\ 
{\tt PLX}   &Trigonometric parallaxes                              & 16329 \\
{\tt pm}    &Proper motions (from various astrometric catalogues)  &243065 \\
{\tt pos}   &Positions      (from various astrometric catalogues)  &668953 \\
{\tt ROT}   &Rotational velocities ($V  \sin i$)                   &  7181 \\
{\tt RVEL}  &Radial velocities of extragalactic objects            & 36552 \\
{\tt SAO}   &Positions and proper motions from SAO catalogue       &252384 \\
{\tt TD1}   &Ultraviolet magnitudes from {\sl TD1} satellite       & 25972 \\
{\tt UBV}   &Johnson $UBV$ photometry                              &141215 \\
{\tt uvby}  &Str\"omgren $uvby$ photometry                         & 37986 \\
{\tt V*}    &Data related to variable stars                        & 25764 \\
{\tt z}     &Redshifts (of distant galaxies and quasars)	          & 88888 \\
\hline 
\end{tabular}
\label{measurements}
\end{table}

For each data type, one can retrieve individual data with
their bibliographical references, and, when available,
weighted means computed from existing observed values by
specialists in the related field.

When measurements are listed as a result of a {\sc Simbad} 
query, they are
normally preceded by a header providing a very short title
to each listed parameter. 

The important r\^ole now played by the {\sc VizieR} database of catalogues
(Ochsenbein et al.\ \cite{vizier}), coming with  easier interoperability
of services, is changing the strategy for inclusion of observational
measurements into {\sc Simbad}.  Let us take the example of
the Hipparcos and Tycho catalogues (ESA \cite{tyc}): once the HIP
or TYC identifier is available from {\sc Simbad} it appears convenient 
enough to  provide the user with a WWW link to the corresponding data in
VizieR rather than overloading the {\sc Simbad} database with the
full Hipparcos and Tycho catalogues.
This functionality is currently being implemented for important
catalogues which have already been cross-identified.
 
As a complement, the WWW interface includes pointers to external archives,
currently:  the INES database of the IUE project
(Rodriguez-Pascual et al.\  \cite{ines}); the high-energy
observational archives at {\sc heasarc} (HEASARC team \cite{heasarc}).

\subsection{Bibliographical references}

One of the key features of the {\sc Simbad} astronomical database
is the unique coverage of bibliographical references to objects.
The bibliographic index contains references to stars from
1950 onwards, and to galaxies and all other objects outside
the solar system from 1983 onwards. Presently 
(November 1999) there are
about 3 million references taken from 110,000 papers
published in the 100 most important astronomical periodical
publications.

\subsubsection{Bibliographical data}

 Articles are scanned in their entirety,  and references
to all objects mentioned in the title, in the abstract,
in the text, in the figures, or in the tables 
are included in the bibliography.
Tables larger than 1000 objects are usually considered as
catalogues and processed separately.

No assessment is made of the relevance of the
citation in terms of astronomical contents: the paper
can be entirely devoted to the object, or simply give a
side mention of it -- in both cases this gives a reference
in {\sc Simbad}.  Note that, for instance, the NED team
(Helou et al.\ \cite{ned})
applies a different strategy when covering bibliography
of extragalactic objects, and tends to select
only  those papers  that appear most relevant. 
Clearly, {\sc Simbad} approach favours exhaustivity,
at the cost of increased information noise.

A code (nicknamed {\em bibcode})
is assigned to each considered paper: 
this 19-digit bibcode contains in principle
enough information to locate the article (including year
of publication, journal, volume, page, etc.). 

When one retrieves the bibliography of a {\sc Simbad} object, a list
of codes is usually given, and -- according to the options used --
these codes are automatically matched against a bibliographic file
which provides the full reference, title 
and list of authors for each citation, 
together with an anchor
pointing to the electronic version of the article.

Currently, in {\sc Simbad}, about 50\% of the objects
have no bibliographic reference.   Among the most cited
objects (more than 2000 references) are the
\object{Large Magellanic Cloud},
\object{M 31},
\object{3C 273}, and the supernova
\object{SN 1987A}.

\subsubsection{Bibliographic reference coding convention}
\label{bibcode}

The structure of the 19-digit \emph{bibcode} 
has been defined in close collaboration with the
NED group at NASA/IPAC so that both databases share the same
coding system (Schmitz et al.\ \cite{bibcode}). 
It is also used, with some adjustments, by the Abstract Service 
of the Astrophysics
Data System (ADS, Kurtz et al.\ \cite{ADS}),
and by the electronic journals (see e.g., Boyce \& Dalterio
\cite{epub}). Reference codes have the following general structure:

\begin{center}
YYYYJJJJJVVVVMPPPPA
\end{center}

\begin{description}
\item[YYYY]
Year of the publication.
\item[JJJJJ]
Standard abbreviation for the periodical.
\item[VVVV]
Volume number (for a journal) or, in the second character
of this field, one of the following
abbreviations for another publication:
  B   (book),
  C   (catalogue),
  P   (preprint),
  R   (report),
  S   (symposium),
  T   (thesis),
  U   (unpublished).
\item[M]
Specific qualifier for a paper:\\
\begin{tabular}{rl}
  L   &  letter  \\
  p   &  pink page  (in MNRAS) \\
  a-z   &  issue number within a volume   \\
  A-K   &  issue designation used by publisher  \\
  Q-Z   &  to distinguish articles on the same page. \\
\end{tabular}
\item[PPPP]
Page number.
\item[A]
First letter of the first author's last name (or `:' if the first
author cannot be identified).
\end{description}

\noindent
Example: 
 {\tt 1991A\&A...246L..24M} \quad for \quad {\sl Astron. Astrophys.} 246,
L24, 1991, a Letter to the Editor of \emph{Astronomy
\& Astrophysics}, by Motch
et al. 

For a  complete description see Schmitz et al. (\cite{bibcode}),
 or the WWW 
server\footnote{http://cdsweb.u-strasbg.fr/simbad/refcode.html}.

\subsubsection{Comments in the references}

Several types of comments are associated with the references
in {\sc Simbad} and normally displayed after the  reference:

\begin{itemize}
\item  General comments: they are often comments
     added by the bibliographers,
     about the problems encountered while cross-identifying
    the objects mentioned in the paper, typos in
    object names, etc.
\item  Notes about the existence of associated electronic
tables, or abstracts in the CDS server.
Papers including no object are also flagged.
\item Information on how the quoted objects are named in
{\sc Simbad} (comments related to the Dictionary of Nomenclature of
Celestial Objects).
\end{itemize}

\subsection{Statistical aspects of the Data Contents}

The astronomical content of {\sc Simbad} results from the complex process
of folding into the database a selection of important
catalogues, and of surveying the complete astronomical literature.

\begin{figure}
\resizebox{\hsize}{!}{\rotatebox{-90}{\includegraphics{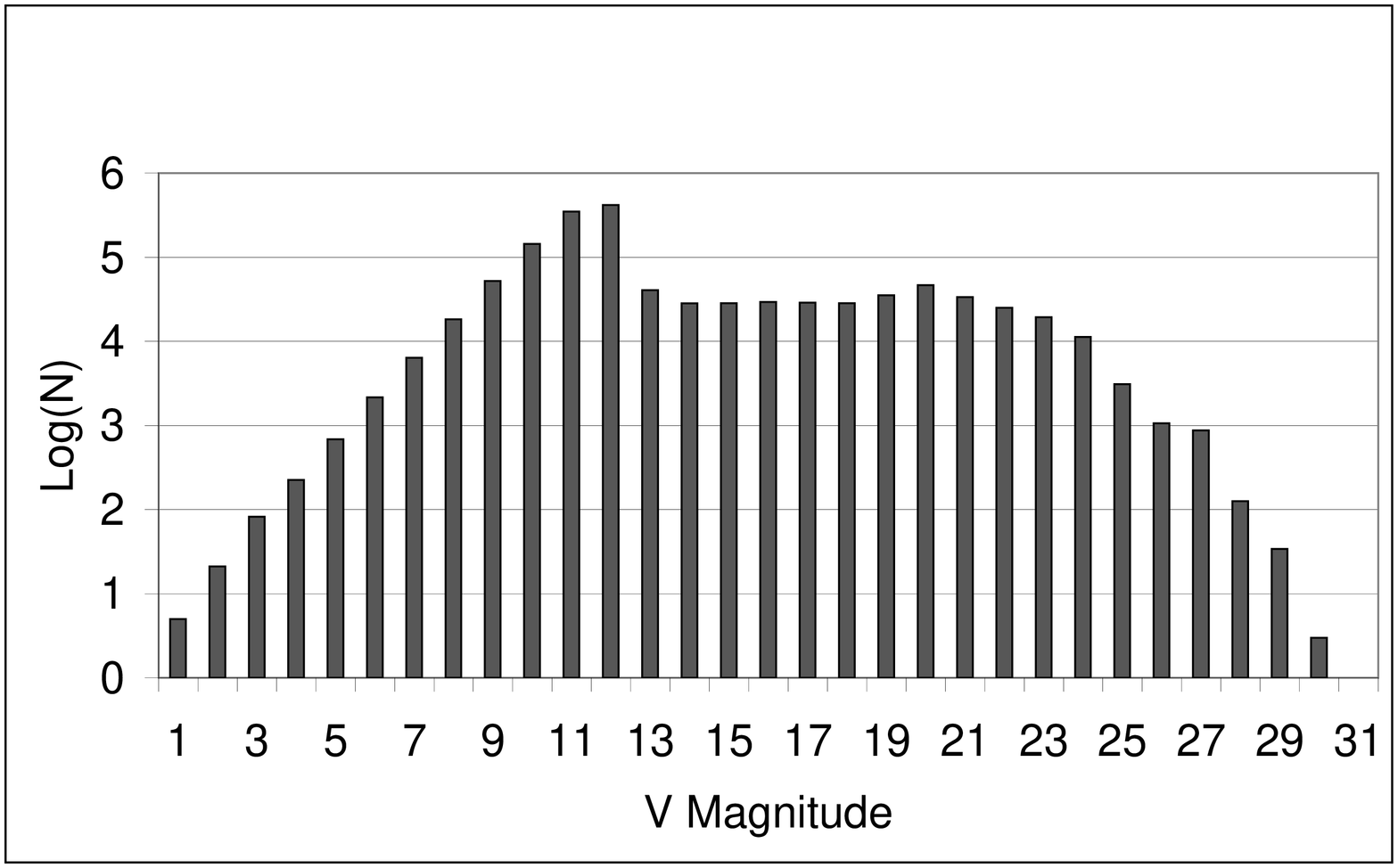}}}
\caption{Histogram of the number of objects in {\sc Simbad} per 
magnitude interval 
(V magnitude; logarithmic scale).}
\label{histo}
\end{figure}

This can be illustrated by the histogram in $V$ magnitudes
of Figure~\ref{histo}. The coverage is
reasonably complete up to beyond magnitude 10 for stars,
after the inclusion of the Tycho catalogue. Many objects
in the range 12 to 26th mag.\ come from extensive studies of objects
in selected sky areas: deep fields,  external galaxies, etc.

Some well-known very large catalogues are not
part of {\sc Simbad}: for instance the Hubble Telescope
Guide Star Catalogue (GSC, Lasker et al.\ \cite{gsc})
is not systematically included (even if GSC identifiers have been added
for all Tycho stars present in {\sc Simbad}).  This
results from a compromise aiming to save database load
as well as manpower for cross-identification and quality
control. Note that {\sc VizieR} and {\sc Aladin} give
access to the full GSC catalogue (and to even larger catalogues
and databases such as USNO-A, DENIS, 2MASS).

\begin{figure}
\resizebox{\hsize}{!}{\includegraphics{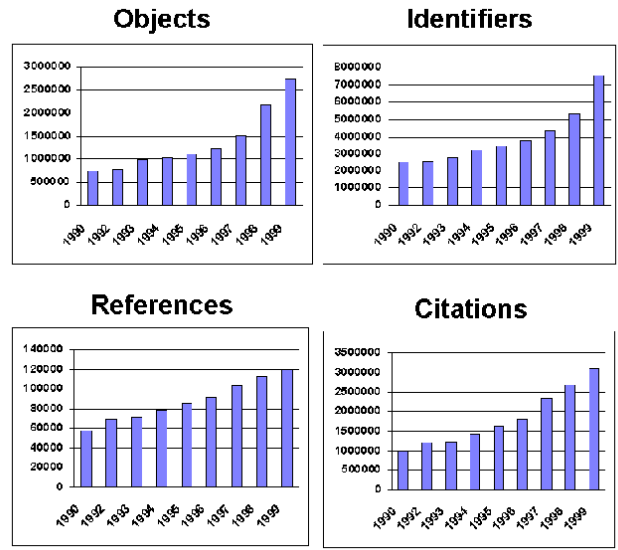}}
\caption{Histograms of different types of database entries for the years
1990 to 1999: number of objects (top left), number of identifiers
(top right), number of references (bottom left), number of citations
(bottom right). The numbers given are the total numbers of entries 
present in the database at the end of the corresponding year.}
\label{graph1a}
\end{figure}

Fig.~\ref{graph1a} illustrates the increase of the
data contents of the database in the years 1990 to 1999.

\section{SIMBAD structure and query management}
\label{dbms}

{\sc Simbad} query mechanism can be summarized by
the following key features:

\begin{itemize}
\item
Database queries can be made mainly through: 
\begin{itemize}
\item  identifiers (names of astronomical objects) and lists of identifiers, 
\item  sets of coordinates (retrieving one object by its position
   on the sky, or extracting all objects lying in a given direction), and 
\item  sampling criteria (or {\em filters\/}).
\end{itemize}
 \item
Data output is driven by formats. The user may
define his/her own formats or modify existing ones.  
Output files can be
saved and mailed to the user.   
 \item
The user interface is adaptable to user
preferences.
\end{itemize}

The database management system of {\sc Simbad} has been developed 
by the CDS, using
the concepts of object-oriented programming. 

\subsection{Object-oriented concepts}

The command language is using the concepts of objects (or agents).
Typical object classes are: astronomical object, object list, database, session,
reference list, filter, format.  Examples of methods are: display, 
describe, bye (quit).

This structure is only visible for the user of the command
line interface. The WWW interface is rendered quite independent
of the database structure. 

\subsection{Indexing}

{\sc Simbad} is organized for optimized access by
identifier (through an index table of object names)
and by position, through an index of small regions. 

\begin{description}
\item[Identifiers]:
A B-tree file contains all identifiers allowing a
fast access to any of them. For each identifier, a record contains
a pointer to the astronomical object itself in the main database.

\item[Position]:
Indexing by coordinates is done in two steps: the coordinates
are mapped into a set of boxes.
{\sc Simbad} uses the spherical-cubic projection
-- a technique also used, e.g., for the Cosmic
Background Explorer (COBE) data: 
the celestial sphere is projected
onto the six faces of a cube, giving six boxes at the first
level. By dividing each face into
four parts, one obtains a partition at level two. Further
levels are obtained by further divisions of each box into four
sub-boxes. Through this mechanism one obtains 6144 boxes at the
level 5 with an average size of 6 square degrees and an average
number of objects of 500. Box {\#}6145 contains all objects
without recorded position.

In order to optimize access to objects in a coordinate box, 
all objects belonging to a box should be physically grouped
in a common place in the database. This is done through a clustering 
mechanism placing objects from the same box in data blocks 
linked together in the database files.
\end{description}

When a set of criteria includes some
limitations in coordinates, this generates the definition of
a list of  boxes including the requested area:
all entries from these boxes are read 
and checked against the whole set of criteria.

When a set of criteria includes no limits in sky position,
the complete database must be scanned -- a long and
somewhat expensive operation, which takes typically
15 minutes in the current hardware configuration.

\subsection{Query by identifier}

In principle any name
found in the literature -- provided it is given as a
syntactically  correct character string -- can be submitted
to the database in order to
retrieve information known for this object.

The general syntax of an identifier is 
the abbreviated catalogue name (or acronym:
generally one to four characters), followed by 
a number or a name (character string) within the catalogue. 

Object names such as Vega and Altair, but also Barnard's star,
Crab Nebula, Sgr A, HDFN, or HDFS 
are stored in the database in a specific catalog 
called `{\sc name}', while star names in constellations, 
such as $\alpha$ Lyrae,
are stored in the catalogue `{\tt *}', 
and variable stars (such as RR Lyrae)
in the catalogue `{\sc var}' (also called `{\tt V*}'). 

The user can generally  type 
{\tt Vega}, {\tt Altair}, {\tt alf Lyrae} (or {\tt alf Lyr}): 
the \emph{sesame} name  resolving module 
(Section~\ref{sesame}) is used for
guessing the catalogue and making the
internal conversion. 
There are however some
difficult cases in which the {\sc name} keyword remains
necessary, such as in {\sc name sgr 1900+14} where
{\sc sgr} stands for Soft Gamma Repeater.

In addition the following hints can help the user
understand the best way to submit an identifier to {\sc Simbad}:

\begin{description}

\item [Case sensitivity:]
{\sc Simbad} is not case-sensitive at this level: {\tt ALF~AQL} or
{\tt alf~Aql} are, for instance, both valid. 
There are some exceptions to the rule, 
such as the cases of the star cluster RMC 136a,
or the star in a multiple system VdBH 25a A,
for which case-sensitivity may be necessary for solving format
ambiguities.

\item [Greek letters:]
should be abbreviated as three letters: {\tt alf}, {\tt bet}, for
$\alpha$ and $\beta$, but also {\tt mu.}  {\tt nu.} and {\tt
pi.} (with a dot), for $\mu$, $\nu$ and $\pi$. 

\item [Constellations:] 
constellation names should be
abbreviated with the usual three letters: {\tt alf Boo}, {\tt
del Sct}, {\tt  FG Sge}, {\tt NOVA Her 1991}.
The full list is
available on-line\footnote{http://simbad.u-strasbg.fr/guide/chB.htx}. 

\item [Multiple systems:]
Identifiers of a multiple system may generate a list of the objects of the
system. For instance, {\tt ADS 5423} calls for the four components, A to D,
of the stellar system around Sirius.  This is true only for
some specific identifiers.

\item[Stellar clusters:]
Clusters which have no NGC or IC number
are named under the generic appellation {\tt Cl}
followed by the cluster name and number: e.g.,
Cl~Blanco~1 is the 1st stellar cluster named by Blanco.
Stars in clusters may belong to a `main' designation
list, or to subsequent lists.
NGC~5272~692 is star 692 in the list by Von Zeipel,
considered as the main list. Subsequent lists have
designations starting with {\tt Cl*}.
Examples: 
Cl*~NGC~5272 AC~968 (list by Auriere \& Cordoni);
Cl*~Melotte~25~VA~13 (13th star in the list by Van
Altena for Melotte 25 -- the Hyades cluster);
Cl* Collinder~110 DI~1101 (list by Dawson \& Ianna -- there
is no `main' list for this cluster).
More details are available in the on-line 
description\footnote{http://simbad.u-strasbg.fr/guide/chC.htx}.

\item [Unknown name ?:]
If the object name seems unknown to {\sc Simbad},
the user is advised to enter the coordinates of the object:
the object may actually exist in the database under a different 
designation. Submitting the identifier, or the name of the
first author of the catalogue, to the
Dictionary of Nomenclature may also give useful clues.

\end{description}

\begin{figure}
\resizebox{\hsize}{!}{\includegraphics{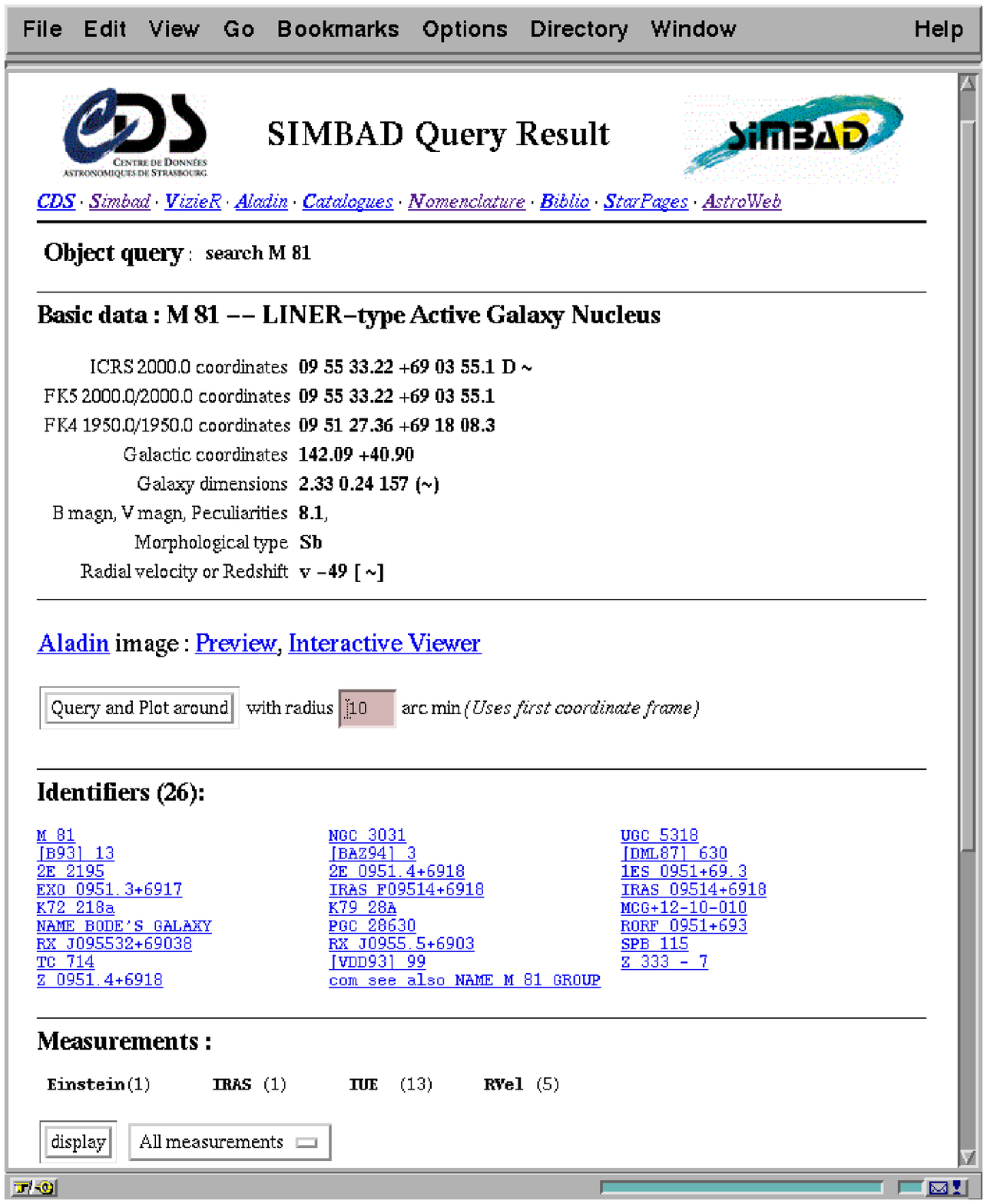}}
\caption{Example of Simbad response page
for a query concerning  M81 (only the first part of the response
is visible here).}
\label{response}
\end{figure}

Figure~\ref{response} illustrates the response received from
the database after submitting the identifier `M~81'.
In the identifier list, the
meaning of acronyms, such as [VDD93], is explained through
a link to the on-line Dictionary of Nomenclature.

The user interface provides an option for querying
around objects, with a radius set by default at 10\arcmin.
This is equivalent to a query by position using the
object coordinates.

It is also possible to generate the list of 10 or 25 next
objects following a given identifier, or to
submit a list of object names,
stored in a file with one identifier per line.

\subsection{Query by coordinates}

Query by coordinates can be used to retrieve all objects 
in a circular field
defined by the coordinates of the center and a radius.

The coordinates can be replaced by the name of
an object lying at the center of the field, in which case
the coordinates are found through an internal query to {\sc
Simbad}. The radius can have any size  (default value is
10\arcmin).  Queries with a radius
smaller than 1--$2\degr$ are answered quite instantaneously.

\subsection{Sampling}
\label{filter}

The sampling mode (also named filter) allows users 
to define criteria for selecting objects in {\sc Simbad}. 

The user may extract objects which satisfy one
set of coordinate criteria, several physical criteria 
(using a simple syntax), objects which have specified 
identifiers or measurements, and, finally, objects having 
citations within a range of years. 

The WWW interface provides an interactive form 
which presents all possible sampling options.

The resulting list may be ordered
according to sort criteria and, furthermore,
it is possible, through the command line mode,
to define precisely the output format.

Note that reading the whole database for extracting a
sample spread on the whole celestial sphere is
possible, but quite time-consuming (as
mentioned above). The user is thus encouraged to
test the filter on a limited region of the sky, before
applying it to the whole database.

\subsection{Charts and sky maps}

After a sampling by position the user can ask for
the corresponding sky plot. This feature is only
available through the WWW interface and is generally optimized
for a radius range of 10--60 minutes.

\begin{figure}
\resizebox{\hsize}{!}{\includegraphics{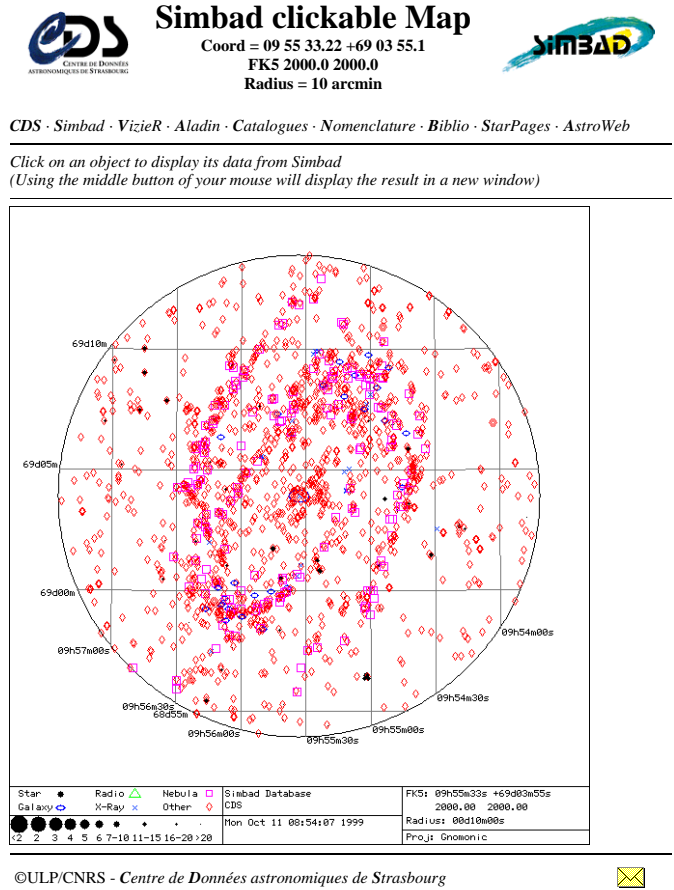}}
\caption{Example of finding chart around M81 (radius of the circular field:
  $10\arcmin$).}
\label{chart}
\end{figure}

The maps display the objects with different
symbols according to object type;  symbol size 
for stars varies with object
magnitude (see Figure~\ref{chart}). The maps are clickable 
and return the object in Simbad corresponding to
cursor position. 

\begin{figure}
\resizebox{\hsize}{!}{\rotatebox{+90}{\includegraphics{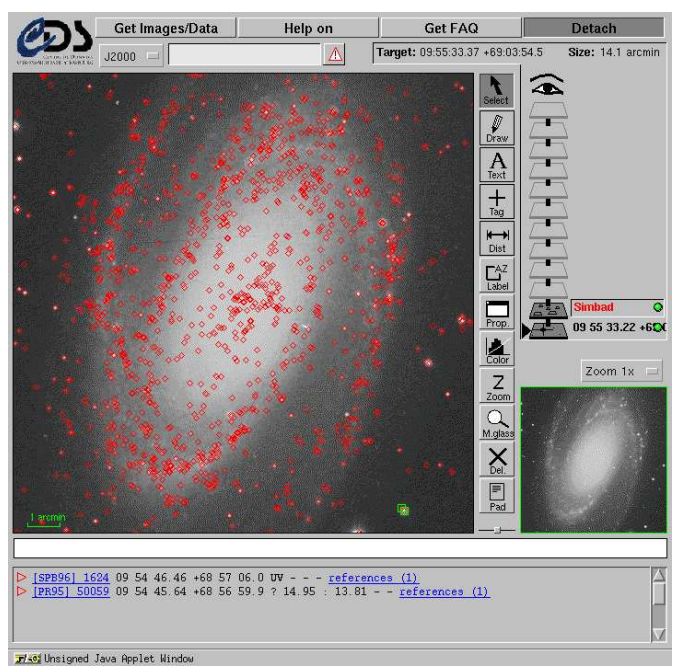}}}
\caption{Use of {\sc Aladin} for displaying {\sc Simbad} entries
  (red diamonds) on top of a DSS-I/STScI image around M81 (width
of the field: $14.1\arcmin$).}
\label{figaladin}
\end{figure}

The WWW interface provides also direct
access to the {\sc Aladin} interactive digitized atlas 
(Bonnarel et al.\ \cite{aladin}) as illustrated in Figure~\ref{figaladin}.

\subsection{Batch mode}

{\sc Simbad} can be queried in {\em batch mode}, by
submitting a mail to the special address
{\tt smbmail@simbad.u-strasbg.fr}.

This is especially useful in case of poor
interactive connectivity, or for submitting 
time-consuming queries or lists.
A WWW form\footnote{http://cdsweb.u-strasbg.fr/simbad/batch.html}
helps to prepare the submission.

\subsection{Resolving a bibliographical reference code}

It is possible to obtain a complete bibliographical
reference, by entering the corresponding
reference code (bibcode).

A reference code can be supplied 
without indicating all
the fields: the first reference corresponding to the
truncated code will be displayed. 
An ampersand ({\tt \&}) should be added
at the end of the truncated bibcode.

\subsection{Additional tools}

Additional tools include special commands for querying
auxiliary databases, on-line help, log files, etc.
More details can be found in the {\sc Simbad} User's Guide
or on the Web pages.

\section{User Interfaces to SIMBAD}
\label{xsimbad}
\label{interface}

There are several user interfaces to {\sc Simbad}.
New users are advised to go directly to the WWW interface,
unless they have very specific needs.

\begin{itemize}
\item The Web interface is currently the easiest
access mode to {\sc Simbad}. This interface takes benefit of the
WWW features to provide the user with additional links
to internal documentation, associated services ({\sc Aladin},
{\sc VizieR}), and external archives (currently: 
the INES database of the IUE project; the high-energy
observational archives
at {\sc heasarc}).  Some features, such as the finding chart,
have been specifically designed for the Web and are 
not available through the other modes.

\item  The command line interface is the basic underlying
interface to the database, which serves as a basis for
 the other more user-friendly interfaces. During many years
it was the sole access mode
to the database, and many users who are accustomed
to the commands may find it quicker and more versatile.
It implies a {\em remote login} (through telnet)
on the {\tt simbad}
machine in Strasbourg, and the user needs to have
a user name and password (see Section~\ref{charge}).

\item  A graphical interactive user interface to {\sc Simbad},
{\sc XSimbad}, 
taking benefit of the X Window environment has been developed 
in 1993-94 for distribution
to  users working in a Unix environment. 
It is now obsolete, because all the functionalities,
and additional ones, are more easily 
available through the Web. 

\item A client/server package is distributed 
on request to data managers
of archives and information systems, when they need to
organize the most efficient access to {\sc Simbad} for
the resolution of object names into the corresponding position,
or the retrieval of other information provided by
the database, such as the reference list for a given object.  
Written currently in C language, it can easily be plugged
into any application able to access C routines.
Distribution is subject to CDS approval.
\end{itemize}

\section{SIMBAD usage}

\subsection{Charging policy}
\label{charge}

{\sc Simbad} is a charged service. The {\tt telnet}
access is protected by userid/password, and the WWW
access is protected either by IP address or by
password. 

Users  have to
register, and get a userid/password from the
CDS staff (or from the U.S. agent for American users).

In the U.S. there is no invoicing for the end-user
because the charges are covered globally 
by NASA for all U.S. users.

In Europe, the same situation is also true for
users from ESO and ESA member states, thanks to an
agreement signed with 
European Southern Observatory and 
ESA Space Telescope European Coordinating
Facility (since January 1995).

Special arrangements also exist or are currently 
being negotiated
with other countries, including Canada, Australia, and Japan.

\subsection{Usage statistics}

There are currently (November 1999) some 7000 user
accounts from 64 different countries.
The development of the WWW access makes difficult 
to keep precise track of the individual usage statistics, 
but the global statistics show that the world wide 
interest for accessing the {\sc Simbad} database continues 
to increase regularly over the years.

The number of {\sc Simbad} queries evolved 
from about 30,000 per month in 1997
to about 100,000 per month in 1999.
About 50\% of the queries come through
the client/server mode.

A mirror copy of {\sc Simbad} has been established 
at CfA (Harvard) for convenience
of US users, and about 
one third of the queries are currently processed on
the mirror site (including name resolving activities for 
the ADS and major US NASA archives).
The mirror copy is managed by CDS and updated
every night.

\section{SIMBAD updates and quality control}
\label{update}

\subsection{Updating SIMBAD}

 {\sc Simbad} is kept up-to-date on a daily basis, as the
result of the collaboration of the CDS team, in Strasbourg,
with bibliographers
in  
Observatoire de Paris (DASGAL), 
Institut d'Astrophysique de Paris, 
and 
Observatoire de Bordeaux 
(Lalo\"e et al.\ \cite{maj93}; Lalo\"e \cite{maj95})
who systematically scan the articles published in some 100
astronomy journals. 

The references are updated very soon after reception of the
journal issues, and in some cases directly from the journal
table of contents, through agreements with the Editors.
New data concerning the objects (identifiers, 
basic data), and new acronyms for catalogs
or tables are being entered
when appropriate.

The inclusion of a large
catalogue in the database is often a long-term task which may
span over several months or
years; the collaboration of specialists in the different fields
is systematically sought.

The improvement of the {\sc Simbad} astronomical contents
relies on a network of collaborations: a list of the
main current contributors is given in  Table~\ref{collaboration}.
More generally, help of other contributing institutes 
and authors, too numerous to be cited here, 
is gratefully acknowledged.
 
\begin{table}
\caption{Main institutes associated to the CDS
for improving the data contents of {\sc Simbad}}
\begin{tabular}{lp{5.8cm}}
\hline 
Bibliography & Observatoire de Paris, 
       Institut d'Astrophysique de Paris, 
       and Observatoire de Bordeaux \\
Astron. contents & GRAAL, Montpellier \\
Galaxies   & Observatoire Midi-Pyr\'en\'ees and
           NASA/IPAC Extragalactic Database \\
Photometry & Observatoire de Gen\`eve
and Institut d'Astronomie de Lausanne \\
Astrometry & Astronomisches Rechen Institut, Heidelberg \\ 
Binary stars & Observatoire de Besan\c{c}on \\
High-energy  & Observatoire de Strasbourg \\
\hline
\end{tabular}
\label{collaboration}
\end{table}

\subsection{Quality control}

 The data contained in {\sc Simbad} are also 
permanently updated, as a result of errata, remarks 
from the bibliographers (during the scanning of the 
literature), integration of lists
and catalogues, quality controls, or special
efforts  initiated by the 
CDS team to better cover some specific domains (e.g., 
multi-wavelength emitters and complex objects). 

Requests for corrections, errata, or suggestions are 
regularly received from {\sc Simbad} users through a 
dedicated {\em hot line}, at e-mail address
{\tt question@simbad.u-strasbg.fr}.
A few dozens of messages are usually received every
week, and processed on a daily basis by the member of
the team who is on duty for that week, or transmitted
to the key person in case of specialized questions.
Remarks received from the users by this way are especially
welcome, as they help the CDS team to improve the database
contents through the scrutiny of specialists' eyes.

Developing new tools for quality control
of the database is a major challenge for the future,
and CDS is exploring possible solutions.
Multivariate analysis applied to
bibliographic information retrieval has
been proven a possible tool for developing quality control
in a database such as {\sc Simbad} (Lesteven \cite{lesteven}).

\subsection{Towards automation of updating procedures}

The advent of 
electronic publishing brings new
perspectives for improvement and automation of the updating
procedures. 

In a first place, tables of contents of the major journals are
now received electronically through the network, 
thanks to journal Editors and Publishers, 
thus reducing the risk of errors.  
Regularly, a number of electronic lists of objects are also 
folded into {\sc Simbad}
through semi-automatic procedures.
 The next step will be the
automatic flagging of object names in the text of
the articles: this has now become 
a very interesting medium-term goal.

Two ways of achieving this flagging are 
currently being considered:
\begin{itemize}
\item the first one is to ask the authors, with
the help of the Editors of electronic journals, to flag
astronomical object names in their text; this can be done, 
for instance, by the use of a \verb+\object{ }+ command within the {\TeX} or
{\LaTeX} source, which will be eventually used to build
an anchor pointing towards {\sc Simbad}, or another
database, in the on-line version made available
on the network.  This approach has been adopted
by the Editors of {\em Astronomy \& Astrophysics}.
\item another approach is the use of intelligent
search tools for identifying object names within the electronic
version of the paper, using a set of syntactic and semantic
rules, and the Dictionary of Nomenclature as a reference
database for already known objects.
\end{itemize}

The first approach seems safer, provided the authors 
understand what exactly they are being required,
and accept this (minor) additional work load.
The latter implies a lot of fine tuning from
the system developers.
The current experience with the handling of publications
(Lesteven et al.\ \cite{lesteven2})
suggests that both approaches may be needed, and that a careful
quality control, including final check by an expert, will
probably remain necessary to avoid errors or misinterpretations,
and to ensure appropriate completeness.

\section{Nomenclature}
\label{Dic}

\subsection{The Dictionary of Nomenclature}
Designations of astronomical objects are often confusing.
A complete list of astronomical designations 
has been collected and published by Lortet et al.\
(\cite{dic2})  in the
{\sl Dictionary of Nomenclature of Celestial Objects 
outside the Solar System}.

This information is available on-line through the {\tt info}
command, or on the 
WWW\footnote{http://vizier.u-strasbg.fr/cgi-bin/Dic-Simbad}.
This service is the electronic look-up version of the 
{\sl Dictionary} which is now under the responsibility
of CDS. It is kept up-to-date on a weekly
basis; about 15 new acronyms
are incorporated every week.

The {\sl Dictionary} currently provides full references 
and usages about some 5000 different acronyms. 
It is used by the International
Astronomical Union as a reference for its recommendations
related to nomenclature.

\subsection{The \emph{sesame} module}
\label{sesame}

The \emph{sesame} module
is used inside {\sc Simbad}
for the management of possible variations in the
naming of astronomical objects.
It is based on a list of rules, written as regular expressions, 
allowing translation of the submitted name
into its {\sc Simbad} canonical form; it is only made visible to
the user when a message mentions the submitted syntax
and its translation.

There are cases where ambiguities cannot be solved.
This is actually specific to the broad context of {\sc Simbad}.
Let us give an example: in the context of extragalactic
objects `N' is a possible abbreviation for `NGC'
(accepted by NED); 
but people studying Novae would frequently use
`N' as an abbreviation for Nova, people studying
\ion{H}{ii} regions would use it for naming
nebulae in the Magellanic Clouds (LHA 120-N or LHA 115-N),
and `N' has also been found in the literature for cluster
stars studied by Nordlund in NGC 2099 (Cl* NGC 2099 N),
for stars studied by Neckel ([N78]), or even for `New'
parts of the galaxy NGC 1275 ([NJS93] in {\sc Simbad}).  
When a name like `N~1992' is submitted
to {\sc Simbad} the ambiguity cannot be solved without
requesting additional information from the user.

\section{Integration of SIMBAD into the CDS Hub}

While the CDS databases have followed different 
development paths, the
need to build a transparent access 
to the whole set of CDS services has become
more and more obvious with the easy
navigation permitted by hypertext tools
(Genova et al.\  \cite{cds2000}). 
{\sc Aladin} has become the prototype of such a development,
by giving comprehensive simultaneous 
access to {\sc Simbad}, 
the {\sc VizieR} Catalogue service, 
and to external databases such as NED,
using a client/server approach and, when possible,
standardized query syntax and formats. 

In order to be able to go further, the
CDS has built a general data
exchange model, taking into account all types of information available
at the Data Center, known under the acronym
of GLU for G\'en\'erateur de Liens Uniformes -- Uniform
Link Generator (Fernique et al.\ \cite{glu}). 

In the current stage of development, the WWW interface
to {\sc Simbad} provides access to  {\sc Aladin}
previewer (reference image around one object),
and to the {\sc Aladin} interactive Java program
(see Bonnarel et al.\ \cite{aladin}).
There are also links between {\sc Simbad} and the
bibliographic services developed or mirrored
at CDS, and more generally to the ADS and the
electronic journals.

While this article is written stronger links between {\sc Simbad} and
{\sc VizieR} are just being created allowing  even easier transfers 
of data and information between both services.
This will also make easier to build new links pointing to
distributed data archives, beyond those already existing
(currently: IUE/INES and HEASARC).


\section{Future developments}

In the near future, the CDS team expects to go on enriching 
the database contents and system functionality.
The users play an important role in
that respect, by giving feedback on the desired features,
on the user-friendliness of the interfaces, etc.

In the context of interoperability of distributed services, 
as currently discussed within the ISAIA project
(Interoperable Systems for Archival
Information Access; Hanisch \cite{isaia}),
{\sc Simbad} is prepared to deliver 
resource profiles and to format the query outputs
in a standard way, for instance XML
 (Ochsenbein et al. \cite{adass9}).

As larger and larger astronomical datasets are being
produced, the CDS is studying the concepts of a new
generation database of several billion objects,
instead of the current several million objects.
We expect {\sc Simbad} to remain an essential
tool for astronomical research in the years to come.

\begin{acknowledgements}
CDS acknowledges the support of INSU-CNRS, the Centre National
d'Etudes Spatiales (CNES), and Universit\'e Louis Pasteur (ULP,
Strasbourg).
Many of the current developments of {\sc Simbad} have been
made possible by long-term support from NASA, ESA, and ESO.
We thank more specifically J.\ Mead and G.\ Riegler (NASA),
P.\ Benvenuti (ESA/ST-ECF),
and P.\ Quinn (ESO) for their help in setting up
the current agreements.

Developing and maintaining the database is a collective undertaking
to which many contributors -- too numerous to be listed here --
are associated. A special mention shall be made of
M.-J.\ Wagner, F.\ Woelfel, J.\ Marcout (Strasbourg), 
A.\ Beyneix, G.\ Chassagnard (IAP, Paris), 
N.\ Ralite,  S.\ Pasquier (Bordeaux), 
E.\ Davoust (Toulouse),
and B.\ Skiff (Lowell Observatory),
who are watching with great care over the {\sc Simbad}
contents.

We want finally to thank Jean Delhaye, Jean Jung, Carlos Jaschek and
Michel Cr\'ez\'e for their leadership and their vision
in the consecutive early phases of  the {\sc Simbad} project.

\end{acknowledgements}


\end{document}